# Single-charge devices with ultrasmall Nb/AlO$_x$/Nb trilayer Josephson junctions

R. Dolata, H. Scherer, A. B. Zorin, and J. Niemeyer

*Physikalisch-Technische Bundesanstalt, Bundesallee 100, 38116 Braunschweig, Germany*

**ABSTRACT**

Josephson junction transistors and 50-junction arrays with linear junction dimensions from 200 nm down to 70 nm were fabricated from standard Nb/AlO$_x$/Nb trilayers. The fabrication process includes electron beam lithography, dry etching, anodization, and planarization by chemical-mechanical polishing. The samples were characterized at temperatures down to 25 mK. In general, all junctions are of high quality and their *I-U* characteristics show low leakage currents and high superconducting energy gap values of $\Delta \approx 1.35$ meV. The characteristics of the transistors and arrays exhibit some features in the subgap area, associated with tunneling of Cooper pairs, quasiparticles and their combinations due to the redistribution of the bias voltage between the junctions. Total island capacitances of the transistor samples ranged from 1.5 fF to 4 fF, depending on the junction sizes. Devices made of junctions with linear dimensions below 100 nm by 100 nm demonstrate a remarkable single-electron behavior in both superconducting and normal state. We also investigated the area dependence of the junction capacitances for transistor and array samples.

## I. INTRODUCTION

Superconducting circuits with small tunnel junctions offer the possibility of manipulating single Cooper pairs and may be used to construct e.g. quantum current standards[1] and charge qubits.[2] For many devices, the use of a material with a larger superconducting gap energy $\Delta$ - compared to the commonly used Al-based devices ($\Delta_{Al} \approx 0.2$ meV) - appears to be attractive. With the use of Nb-based junctions ($\Delta_{Nb} \approx 1.5$ meV) it should be possible to increase the Josephson coupling energy $E_J$ by almost one order of magnitude for the same tunnel resistance (or transparency) of a tunnel barrier. This opens a direct way towards the realization of the required relation between $E_J$ and the charging energy $E_C$ - namely $E_J \approx E_C \gg k_B T$, where $k_B T$ is the thermal energy - in very small Nb circuits. Owing to the fact that both parameters, $E_J$ and $E_C$, are large compared with the thermal energy, such circuits are particularly promising for the development of sensitive single Cooper pair electrometers[3] and charge-phase qubits.[4]

Traditionally, single Cooper pair and single electron tunneling (SET) devices are fabricated by angle evaporation techniques using suspended masks, defined by high-resolution electron beam lithography (EBL). This technique works successfully for soft materials like aluminum which is the most widely used metal for the preparation of SET devices. Attempts to use this technique for Nb-based devices resulted in a deterioration of the superconducting properties of the Nb; the transition temperature $T_c$, and hence the effective gap energy, is reduced to about 1 meV or smaller[5,6,7,8], i.e. significantly below the Nb bulk value. Furthermore, appreciable leakage currents were observed in the subgap region.[8] Nb/AlO$_x$/Nb junctions with areas as small as 0.002 µm$^2$, good $I$-$U$ characteristics and a gap energy of about 1.25 meV have been achieved by a sloped-edge technique.[9] Unfortunately only single junctions have been presented so far whereas more complex devices might be difficult to fabricate by this method.

For applications requiring larger junctions with linear dimensions above 1 µm, e.g. superconducting digital electronics,[10] the junctions are normally prepared from Nb/AlO$_x$/Nb trilayers.[11] Trilayers are deposited *in situ* on the whole wafer and patterned by etching or anodization techniques. They provide the best $I$-$U$ characteristics with low subgap leakage currents and a high gap energy, approaching the bulk value. So far, only few results have been published about SET effects in devices prepared from trilayers.[12,13,14] Early works used standard etching techniques and planarization by chemical-mechanical polishing (CMP)[12] and spin-on-glass (SOG).[13] These devices still suffered from rather high capacitances due to large junction areas and showed small gate modulation amplitudes. Recently good progress has been achieved by using an ingenious focused ion beam etching technique to prepare a single-electron transistor with small junction areas from a Nb/AlO$_x$/Nb trilayer.[14] This work demonstrates the potential of Nb devices, however this fabrication technique, based on the processing of individual junctions one by one, is time-consuming and practically limited to devices and circuits containing only a small number of junctions. For future medium- or large-scale integration, the standard lithographic and etching techniques appear to be more favorable.

We developed an optimized fabrication process using EBL, dry etching, anodization, and CMP planarization, which allows the preparation of small Josephson junctions starting from standard Nb/AlO$_x$/Nb trilayers. By means of this process, high-quality junctions with dimensions below 100 nm × 100 nm can be fabricated with a high degree of reproducibility, enabling at least medium-scale integration. The process is partly based on previously published works.[13, 15, 16]

## II. SAMPLE FABRICATION

Although modern optical lithography is able to create patterns below 100 nm, all layers are defined by EBL in our process. Particularly with regard to frequent layout changes for process development and low volume production, EBL is much more flexible as no expensive photomasks are needed. The fabrication process flow is shown in Fig. 1. Standard Nb/AlO$_x$/Nb trilayers are deposited on oxidized three-inch silicon wafers. At first a Nb base layer, 150 nm in thickness, is deposited by dc-sputtering and covered by an rf-sputtered Al layer 10 nm in thickness. The barrier is created by oxidation in pure O$_2$ at a pressure of 500 Pa for 10 minutes at a temperature of about 20 °C. Thereupon the Nb counterelectrode, also 150 nm in thickness, is deposited. On top of the trilayer a thin SiO$_2$ layer, 100 nm in thickness, is deposited by plasma enhanced chemical vapor deposition (PECVD). This layer was found to improve the reproducibility of the first etching process and serves as a mask in a forthcoming anodization process. Using EBL and a two-layer polymethylmethacrylate (PMMA) resist, the mask for the first etching process is created by lift-off of a thermally evaporated Al layer, 30 nm in thickness. This Al mask defines the area of the Josephson junctions. The SiO$_2$ layer and the Nb counterelectrode are patterned one after the other by reactive ion etching (RIE) in CHF$_3$ and CF$_4$. The etching parameters have been carefully optimized to ensure a good pattern transfer from the Al mask to the Nb counterelectrode with minimal size deviations. Usually a slight increase in area of the junctions compared with the mask is observed in our process. Since the CF$_4$ RIE process does not etch Al, the AlO$_x$/Al barrier and the Al mask layers are removed by a short Ar ion beam milling step. Then the wafer is anodized up to a voltage of 20 V, yielding a Nb$_2$O$_5$ layer thickness of about 45 nm. This anodization step was found to decrease subgap leakage currents. Leakage currents were observed in several non-anodized samples and their magnitude was related to the etching parameters. They are probably induced by damage of the junction perimeter or by re-deposition of conductive material due to the etching process. The anodization also decreases the areas of the Josephson junctions[14]. By proper selection of the anodization voltage, the increase in area caused by the etching process can be compensated.

By use of a second EBL process, now in combination with a negative tone resist, a mask for patterning the base electrode is defined. Then the Nb$_2$O$_5$/Nb layer is etched in an SF$_6$ RIE process. After removing the resist, a PECVD SiO$_2$ layer of 500 nm thickness is deposited. This layer is planarized by CMP until the Nb counterelectrode is exposed at the surface. Since the optimum polishing time for each device on the wafer depends on the junction sizes and local pattern density, large dummy junctions are placed around the active junctions to ensure the same clearing time, independent of the size of the active junctions. Finally, after thorough Ar plasma cleaning of the samples, a 150 nm thick Nb wiring layer is deposited by sputtering and patterned by RIE in CF$_4$ through a third, EBL-defined Al mask.

Using the described process, several wafers have been fabricated containing single electron transistors with different nominal junction areas $A_{nom}$ ranging from 80 nm × 80 nm to 180 nm × 180 nm, arrays consisting of 50 small junctions in series and large junctions of 3 µm × 3 µm size as a reference for the determination of the trilayer parameters. Fig. 2 shows a micrograph of a typical transistor.

## III. MEASUREMENTS AND RESULTS

Prior to their low-temperature investigation, the resistance of the samples was inspected at room temperature. Some samples were checked for their properties at $T = 4.2$ K by measuring the *I-U* characteristics in a liquid helium bath. On some wafers a considerable fraction of the transistors were found to be shorted, in marked contrast to good *I-U* curves for the junction arrays and the large junctions. These defects are probably due to static discharge and not to the fabrication process itself.[16]

Selected chips were investigated in a dilution refrigerator at a base temperature of $T = 25$ mK. Electrical biasing and signal amplification were performed using symmetrical voltage (or current) bias and low-noise preamplifiers with all signal lines rf filtered by Thermocoax™ cable pieces 1 m long. A magnetic induction of several Tesla could be applied using a solenoid coil to quench the superconductivity of the samples. The main parameters of the transistor samples with junction sizes nominally ranging from 80 nm × 80 nm to 150 nm × 150 nm are given in Table I.

As a global inspection demonstrates (Fig. 3), the *I-U* curves of the samples in the superconducting state were found to be equal in shape, showing small subgap leakage and sharp onset of current at the gap voltage. The transistor's gap voltage $U_{gap} = 4\Delta/e$ was determined from the maximum slope in the *I-U* curves (see the derivatives plot in Fig. 3b), where a slight dependence of $U_{gap}$ on the junction area size appeared, i.e. $U_{gap} = 5.3$ meV for the smallest transistor and 5.5 meV for the largest one. The corresponding energy gap of $\Delta \approx 1.33$ meV - 1.38 meV is typical for all trilayers fabricated in our lab and can be considered as a characteristic of the present Nb target, since this value was also found in larger junctions prepared with a different fabrication process. Table I also shows the nominal junction areas $A_{nom}$, defined in the sample layout, in comparison with the experimentally evaluated values $A_{exp}$. The latter values were derived by comparing the normal-state resistance $R_\Sigma$ of the small double-junction samples with a large 3 µm x 3 µm reference junction. We found that $A_{exp}$ was slightly larger than $A_{nom}$ (relative deviations 10% – 15%) except for the smallest sample T1, where $A_{exp} \approx 70$ nm × 70 nm appeared 20% smaller than $A_{nom}$. These deviations can be attributed to a not completely optimized proximity effect correction in the EBL and possibly to size-dependent effects in the etching and anodization processes.

*I-U* characteristics of the investigated transistors also show small supercurrent features at zero bias which are visible in Fig. 3. This feature is clearly seen as an antisymmetric current peak even in the smallest transistor (T1) whose *I-U* curve is zoomed in Fig. 4. Below we will first focus on this sample because it shows a periodic dependence of this feature on the gate voltage. The observed height of the supercurrent peak, shown in Fig. 4, i.e. $I_p \approx 60$ pA, is much smaller than the Ambegaokar-Baratoff value of the Josephson critical current in individual junctions, which is equal to $I_c = 28$ nA.[17] (Note that the measurement of this sample in another, four-terminal current biased configuration with ramping the bias current at a rate $dI_{bias}/dt \approx 30$ pA/s gave the value of switching current $I_{sw}$ close to $I_p$, i.e. $I_{sw} \approx 0.15$ nA, see Fig. 4). Taking into account a possible reduction of the transistor supercurrent due to the charging effect on the island (for the evaluated parameters $E_J \approx E_C$), the expected reduction of the critical current in this transistor can hardly result in a value smaller than $0.2I_c \approx 6$ nA (see, for example, the plot in Fig. 2a of Ref. 18). Therefore, a significant (about two orders of magnitude) suppression of the supercurrent in this experiment can be attributed to a strong effect of noise. In this regime fluctuations of the Josephson phase across the transistor are very large, so the average (observable) supercurrent is suppressed dramatically. Similar behavior was observed in Al devices (see, for example, Refs. 19-22). The supercurrent peak shape is determined by the value and frequency dependence of the real part of the electromagnetic impedance of environment (including the self-capacitance of the Josephson element) in a wide frequency range as well as by the effective temperature. For the given voltage bias configuration the low-frequency impedance of the lines is about 1.6 k$\Omega$, while at radio and microwave frequencies its value is of the order of 100 $\Omega$. Small supercurrent peak magnitudes can be treated as a result of the regime of stochastic tunneling of individual Cooper pairs, which is described by the perturbation theory.[23] This theory gives the values of the observable supercurrent proportional to the square of the Josephson coupling strength. In the case of transistors, such tunneling of pairs occurs through the double junction, so one can say that this is a Cooper pair cotunneling effect. The effective Josephson coupling strength of the transistor $E_{tr}(V_g)$ periodically depends on the gate voltage,[18, 24] and for the ratio $E_J/E_C$ less than or about unity it takes a simple analytical form [25]

$$E_{tr} = \frac{E_J^2}{4E_c}\left[1-(Q_0/e)^2\right]^{-1}, \qquad Q_0 = C_g V_g \bmod(2e) \neq e \bmod(2e), \qquad (1)$$

where $C_g$ is the gate capacitance. The dependence of the observable supercurrent on the gate voltage should in the given regime of operation mimic the dependence $[E_{tr}(V_g)]^2$. We tried the gate modulation of the transistor current in the superconducting state, voltage-biasing the samples close to zero voltage corresponding to the maximum of the supercurrent peak. On the smallest transistor T1 we found a clear current modulation, showing the shape of a set of characteristic segments (see Fig. 5). The gate voltage period of 15 mV corresponded to $e$-periodic modulation, as a comparison with later measurements in the complete normal state showed (Fig. 6). Thus, because of poisoning of the island by quasiparticles the observed dependence was formed by the lowest segments of two $e$-shifted dependencies (in the island charge range from $-e/2$ to $e/2$) corresponding to even and odd parity states. The modulation amplitude was about 90 pA, corresponding to a peak-to-valley ratio of about 2. The dependence Eq.(1) gives the ratio $E_{tr}(C_g V_g = e)/E_{tr}(C_g V_g = 0) = 4/3$ that leads to a peak-to-valley ratio $(4/3)^2 \approx 1.78$. This value is close to the observed value of 2. Larger values of $E_J/E_C$ (or, in other words, smaller values of $E_C$) should reduce the observed value of the peak-to-valley ratio because of a less pronounced charging effect (cf. Fig. 2a in Ref. 18). Therefore, the observed shape of the $V_g$ dependence of the supercurrent (cf. Fig. 5) confirms our suggestion about Cooper pair tunneling in the presence of large fluctuations and indicates that the capacitance of the island is relatively small. So the charging energy of the transistor is not smaller than the Josephson coupling energy of individual junctions, i.e. $E_C \geq E_J$. This conclusion is in accordance with the parameters for transistor T1 given in Table I. Larger transistors (T2-T4), although exhibiting respectively larger supercurrent peaks (see Fig. 6), did not show notable gate modulation of the supercurrent.

Further, the subgap region of T1 showed a nearly linear current rise starting at about 2.5 mV, i.e. at about half of the gap voltage of the transistor. Assuming an equal division of the voltage between the junctions it corresponds to a voltage across each junction equal to $\Delta/e \approx 1.25$ mV. Although this rise is small in comparison with the onset at $U = 2\Delta/e$, the quasiparticle current in this voltage region is comparable with the supercurrent peak. The property of a noticeable increase in the quasiparticle current at $\Delta/e$ is characteristic for the low-temperature behavior of our tunnel junctions as well as some large Nb-based tunnel junctions fabricated in other labs (see, e.g., Ref. 26) and is likely related to a small number of intragap states in the superconducting Nb electrodes of the junctions. This rise in the quasiparticle current in individual junctions at $U \approx \Delta/e$ leading in combination with a supercurrent branch to an N-shape non-linearity of the junctions seems to be the reason for an additional peak-like feature in the transistor characteristic observed at a voltage around 1.5 mV (see the curve corresponding to the voltage biased mode in Fig. 4). Due to the negative-conductance branches of the non-linear $I$-$U$ characteristics of individual junctions at such voltage bias the total voltage across the transistor, $U = U_1(I) + U_2(I)$, is redistributed unequally between the junctions. Most of the bias voltage drops across one junction, say, junction 2, i.e. $U_2 \approx U \approx \Delta/e$, while only a small voltage is now applied to junction 1 which switches into the supercurrent regime, $U_1 \ll U$. This stable voltage configuration is possible until an increase in total voltage $U$ leads to an increase in both $U_1$ and $U_2$ accompanied by a rise in the current. So, the tunneling of Cooper pairs across junction 1 is completed by the tunneling of quasiparticles across junction 2. The difference of this regime from the one established on the supercurrent peak is that the effective electromagnetic impedance of environment now includes the serially-connected high impedance of junction 2. As a result, the dependence of the transistor current on the bias voltage takes the shape of a peak with a height that is somewhat smaller than the height of the main supercurrent peak. It is remarkable that the top of this peak can be modulated by gate voltage (*e*-periodic modulation, not shown), demonstrating the regime of Josephson-quasiparticle (JQP) cycles in this (rather narrow) voltage range.[27] The observed depth of this modulation in comparison with that of the supercurrent peak was rather small.

A similar peak-like behavior was observed in $I$-$U$ characteristics of larger transistors, except for the JQP modulation of the current. Fig. 7 shows these features for transistors T1-T4 in one plot. One can see that the amplitude of the peak is steadily increased with the junction size. This is clearly related to the fact that the larger junctions having larger Josephson coupling energy can carry a larger supercurrent in this regime. The left wings of the peaks mimic the rise in the quasiparticle current in the transistor, which is reduced to one (having most of the voltage drop) junction. This is seen as an approximately two times larger differential conductance d$I$/d$U$ on this wing than the one observed at two times larger voltage, at which the bias voltage drops equally on the junctions. On the contrary, the right wings of the peaks repeat the shape of the current drop on the right wings of the main supercurrent peak. At larger voltage behind the peak the voltage drop on junction 1 increases towards equaling the individual voltage drops and both junctions predominantly conduct quasiparticles.

We also investigated the two series arrays consisting of $N = 50$ junctions with $A_{\text{nom}} = 100$ nm × 100 nm (sample 50T2) and 150 nm × 150 nm (sample 50T4), prepared on the same chip. Their $I$-$U$ curves are shown in Fig. 8 and their main sample parameters are given in Table II. In contrast to the transistor samples the measured junction sizes $A_{\text{exp}}$ here appeared two times larger than the nominal values $A_{\text{nom}}$. This can probably be attributed to the proximity effect in the EBL, since the same electron dose was used for the exposure, but the environment of the junctions in the arrays is different compared to the transistors. Also, microloading effects in the RIE process cannot be excluded.

The gap feature voltage of 131.3 mV measured on both arrays is consistent with the result obtained on the transistor samples, indicating that all junctions were in order, and corresponds to $2\Delta = 2.63$ meV. The supercurrent through the arrays shows as a characteristic peak at low voltage. In the subgap range of the *I-U* characteristics we found remarkable oscillatory structures, fading out with increased voltage (Fig. 8 b). Period (1.26 mV to 1.27 mV) and shape of the current oscillations were similar for both arrays. Similar to double-junction samples we interpret the sequence of the observed current peaks as a result of the re-distribution of the bias voltage between the junctions of the arrays. For example, the first (after the supercurrent) peak corresponds to the main voltage drop $\geq \Delta/e$ across one junction carrying quasiparticles, while the rest *N*-1 junctions with small individual voltage drops carry supercurrent (conduct Cooper pairs). That single junction plays the role of an additional environmental impedance connected in series with the bias line impedance. A further increase in voltage results in the switching of one more junction in the quasiparticle current regime, so the impedance of two junctions is now connected in series to the chain of *N*-2 junctions carrying supercurrent. Thus the second peak is positioned at a voltage which corresponds to the onset of the quasiparticle current in two junctions, and so forth. Note that a similar oscillating behavior was observed by Haviland *et al.* in long arrays of small Al junctions.[28] The period of fading current oscillations in their experiment was close to 400 µV $\approx 2\Delta_{Al}/e$. Taking into account the fact that the quasiparticle-current onset in Al tunnel junctions starts only at the gap voltage $2\Delta_{Al}/e$ one can understand the reason why the period of the current oscillations in Al arrays is determined by the double-gap voltage.

After completing the sample characterization in the superconducting state we applied a magnetic field to drive the structures to the normal state. It showed that a magnetic inductance $B = 8$ T was necessary to fully quench the superconductivity in the Nb films. Total island capacitance values $C_\Sigma$ for the transistors were determined by using the standard method,[29] i.e. by plotting $(U - I\,dU/dI)$ vs. $U$, deriving the offset voltage $U_{off}$ and calculating $C_\Sigma = e/U_{off}$ (Fig. 9). Values for $C_\Sigma$ ranging from 1.5 fF to 4 fF were found, depending on the junction sizes. Next, the single-junction capacitances were estimated according to $C_T \approx 1/2\, C_\Sigma$, taking into account that the gate capacitance contribution $C_g \approx 0.01$ fF for all samples was negligible.

A second, independent way to determine the junction capacitances is given by the Coulomb blockade thermometer (CBT) method.[30] Here, the *I-U* characteristics are evaluated at higher temperature where the $k_B T$ competes with the charging energy. Under this condition the Coulomb blockade manifests as a bell-shaped dip in the voltage dependence of the differential conductance $G = dI/dU$ around zero bias, as shown in Fig. 10. For the parameter $u = 2((N-1)/N)(e^2/2C_T)/k_B T$ ($\approx 1$ for our measurements at $T = 1.2$ K), the normalized depth of this dip is given by $G_0/G_T \approx 1 - 1/6\, u + 1/60\, u^2 - 1/630\, u^3$, where $G_T$ is the conductance in the high voltage limit and $N$ and $C_T$ are the number and capacitances of the junctions in the homogenous series array. Following this line we derived $C_T$ values for both array and transistor samples. Possible corrections due to rather low $R_T$ (especially for array 50T4) were checked and found to be negligible.[31] We also fitted the normalized conductance curves with corresponding theory functions,[30] which yielded consistent $C_T$ and $T$ values, as shown in Fig. 10 for the array samples.

In Fig. 11 the junction capacitances for the array and transistor samples are plotted against the junction areas. It can be seen that for the transistor samples, where $C_T$ was obtained by the two independent methods (offset plot and CBT), both resulting values agree well within the experimental uncertainties, which again proves their soundness. In Fig. 11 it is striking that the dependence $C_T(A_{exp})$ obeys a linear relation. We therefore fitted $C_T(A_{exp}) = C_0 + c\, A_{exp}$ to the data, yielding $c = (59 \pm 4)$ fF/μm$^2$ and $C_0 = (0.43 \pm 0.05)$ fF. The value of $c \approx 60$ fF/μm$^2$ for the specific junction capacitance is in good agreement with values found in literature for comparable junctions with similar current density,[11, 32] e.g. Ref. 11 gives a value of $c \approx 60$ fF/μm$^2$ and Ref. 32 of $c \approx 50$ fF/μm$^2$. However, the data suggest that a large offset capacitance ≈ 0.4 fF dominates the total island capacitance at very small junction sizes (see dotted line in Fig. 11). Since this becomes a problem when aiming at a further reduction of $C_\Sigma$, we looked for the origin of this offset capacitance, assuming that this adds up from two components: The first one is a parasitic capacitance appearing parallel to the junctions at their circumference, where the Nb$_2$O$_5$ layer with large dielectric constant ($\varepsilon \approx 29$) is present. We estimated this effect using a simple plate-capacitor model and found that, due to the small effective cross section of the Nb$_2$O$_5$ layer, this capacitance contribution is quite small (about 0.05 fF – 0.1 fF). The second and, in fact, dominant contribution is the stray capacitance $C_S$ between the island body and the surrounding source and drain electrodes. Since the geometry of the transistors is rather complex and not exactly known for the real samples (see Figs. 1 and 2, in particular the thickness of the dielectric layer after the CMP process is uncertain), an exact analysis of the island's stray capacitances was not feasible. Therefore, we again applied a simple plate-capacitor model, plugging in the overlapping areas and the distance of island and electrodes. This estimate yielded $C_S \approx 0.15$ fF - 0.30 fF, which in sum with the Nb$_2$O$_5$ layer contribution approaches the offset capacitance $C_0$ evaluated from the data fit. We conclude that elaborated changes in the layout of the samples will be necessary to reduce this stray capacitance considerably.

## IV. CONCLUSION AND OUTLOOK

Standard Nb/AlO$_x$/Nb trilayers have been used to fabricate high-quality mesoscopic Josephson junction transistors and junction arrays. The smallest junction area achieved up to now is 70 nm × 70 nm, leading to a junction capacitance of 0.75 fF and complying with the desired relation between the Josephson coupling energy and the charging energy $E_J \approx E_C = 55$ μeV. The dependence of the total capacitance of the transistor island on the junction area shows quite a large parasitic capacitance of ≈ 0.4 fF, resulting primarily from the overlap between the transistor island and the drain/source electrodes. A reduction of this capacitance can be expected by a factor of about two by proper optimization of the layout and the thickness of the dielectric layer. Preliminary Nb etching tests in a modern etching system equipped with an inductively coupled plasma (ICP) source resulted in improved sub 100 nm patterns with a high degree of reproducibility. By using this process for the definition of the junction areas a further reduction of the junction size and improved run-to-run repeatability can be achieved.
The parameters of the single Cooper pair transistors achieved in this paper are comparable with those typical of the charge-phase qubit samples in which the transistor is either connected to a larger readout junction[4] or closed by a low-inductance superconducting loop.[33] Therefore, our Nb transistors can be tested in a qubit-operation regime. Moreover, using Nb for the qubit element itself makes it, in principle, possible to implement on the same chip a Nb-based readout circuit like, e.g., a radio-frequency tank circuit or a balanced RSFQ comparator.[34] One of the problems of realizing Nb-transistor-based qubits seems to be their imperfection with respect to the *e*-periodic behavior of the dc characteristics, indicating an appreciable effect of quasiparticle poisoning of the island. Further investigations will be necessary for a better understanding and improvement of this property which may appreciably shorten the characteristic coherence times of the qubit.

## ACKNOWLEDGMENTS

The authors would like to thank Th. Weimann for his advice regarding EBL, B. Egeling for his support in the CMP and PECVD processes, K. Schuster for the ICP etching tests, M. Götz and S. A. Bogoslovsky for assistance in the measurements and D. B. Haviland for fruitful discussions. This work was partially supported by the EU within the scope of the SQUBIT-2 project.

**Tables**

TABLE I: Parameters for the transistor samples investigated. $R_\Sigma$ is the normal-state resistance of the series double junction. $A_{nom}$ is the nominal junction area in the sample layout, while $A_{exp}$ is calculated from the relation $R_\Sigma \times A_{nom} = 0.77$ k$\Omega \times \mu m^2$, measured on a large (3 µm × 3 µm) reference junction. The superconducting gap energy $\Delta$ is obtained from the maximum slope in the I-U curves (Fig. 3b). $C_T$ is the capacitance of a single junction, averaged from the values derived from the voltage offset plot and from the CBT method (see text). The Josephson coupling energy $E_J \equiv (h/4\pi e)I_c$ was calculated using the Ambegaokar-Baratoff relation $I_c = (\pi/2e)\Delta/R_T$, with $R_T = R_\Sigma/2$ and $R_K = h/e^2 \approx 25.8$ k$\Omega$. $E_C = e^2/2C_\Sigma$ is the charging energy of the transistor island ($C_\Sigma \approx 2C_T$).

| Transistor sample | $R_\Sigma$ (k$\Omega$) | $A_{nom}$ (µm$^2$) | $A_{exp}$ (µm$^2$) | $2\Delta$ (meV) | $C_T$ (fF) | $E_J$ (meV) | $E_C$ (meV) | $E_J / E_C$ |
|---|---|---|---|---|---|---|---|---|
| T1 | 152.4 | 0.0064 | 0.005 | 2.65 | 0.75 | 0.056 | 0.054 | 1.0 |
| T2 | 72.4 | 0.0100 | 0.011 | 2.70 | 1.1 | 0.120 | 0.036 | 3.3 |
| T3 | 52.7 | 0.0144 | 0.015 | 2.71 | 1.3 | 0.166 | 0.029 | 5.7 |
| T4 | 29.9 | 0.0225 | 0.026 | 2.75 | 1.9 | 0.297 | 0.021 | 14 |

TABLE II: Parameters for the array samples investigated, consisting of 50 junctions in series. $C_T$ is the single junction capacitance, derived from the CBT method. $E_C^{sj} = e^2/2C_T$ is the charging energy related to a single junction. All other parameters are defined and were obtained as explained in Table I.

| Array sample | $R_\Sigma$ (k$\Omega$) | $R_T$ (k$\Omega$) | $A_{nom}$ (µm$^2$) | $A_{exp}$ (µm$^2$) | $2\Delta$ (meV) | $C_T$ (fF) | $E_J$ (meV) | $E_C^{sj}$ (meV) |
|---|---|---|---|---|---|---|---|---|
| 50T2 | 970 | 19.4 | 0.010 | 0.02 | 2.63 | 1.9 | 0.22 | 0.042 |
| 50T4 | 490 | 9.8 | 0.023 | 0.04 | 2.63 | 2.7 | 0.43 | 0.029 |

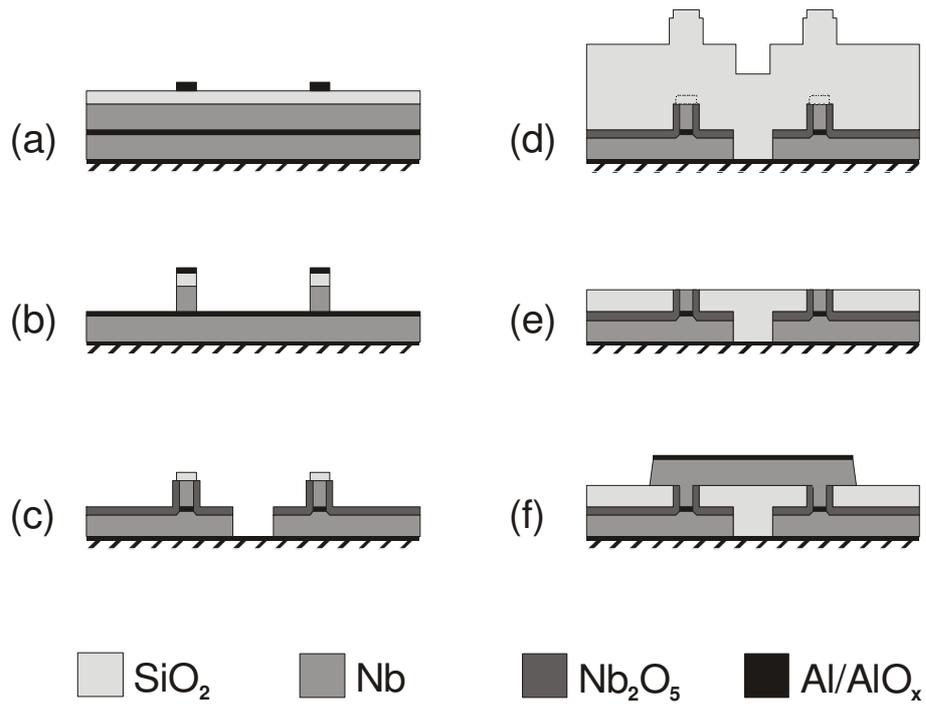

Fig. 1: Main process steps for the fabrication of a Nb/AlO$_x$/Nb trilayer transistor. (a) Al etching mask on trilayer with SiO$_2$ overlayer, (b) etching of SiO$_2$ and top Nb layer, (c) removing of Al, anodization, etching of bottom Nb layer, (d) deposition of SiO$_2$ layer, (e) CMP planarization, (f) deposition and etching of Nb wiring layer.

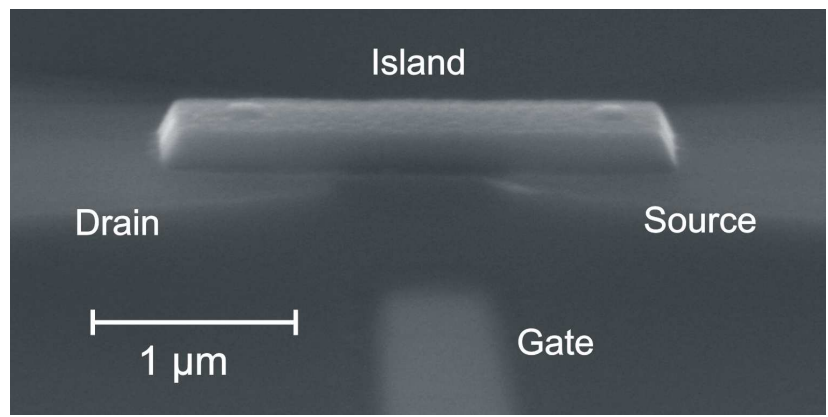

Fig. 2: Scanning electron micrograph of a Nb/AlO$_x$/Nb trilayer transistor. Drain, source and gate are buried in SiO$_2$.

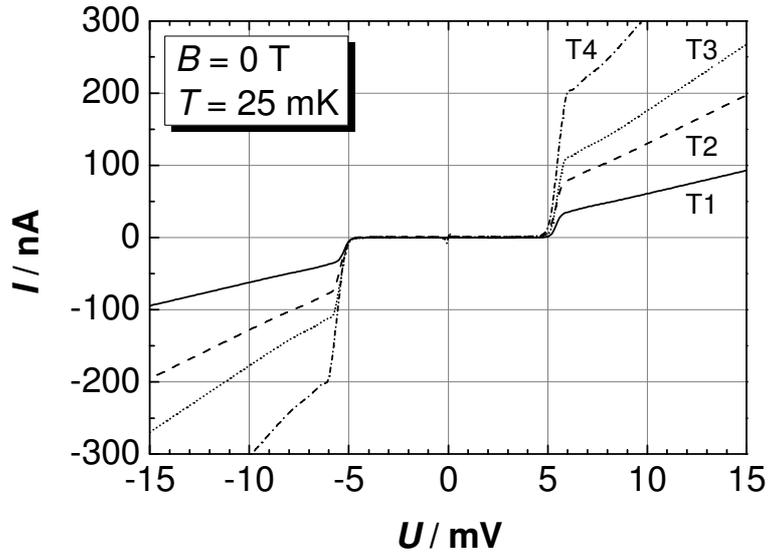

(a)

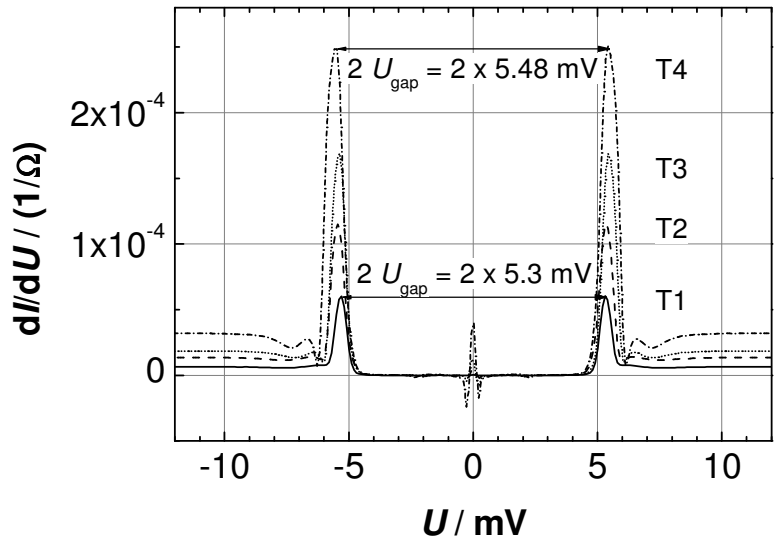

(b)

Fig. 3: *I-U* curves (a) and their derivatives (b) of transistor samples T1-T4, measured in the superconducting state. The peak maxima positions in the d*I*/d*U* curves were identified with the gap voltage $U_{gap} = 4\Delta/e$. A weak sample or junction size dependence of $\Delta$ is apparent (see Table 1 for sample parameters).

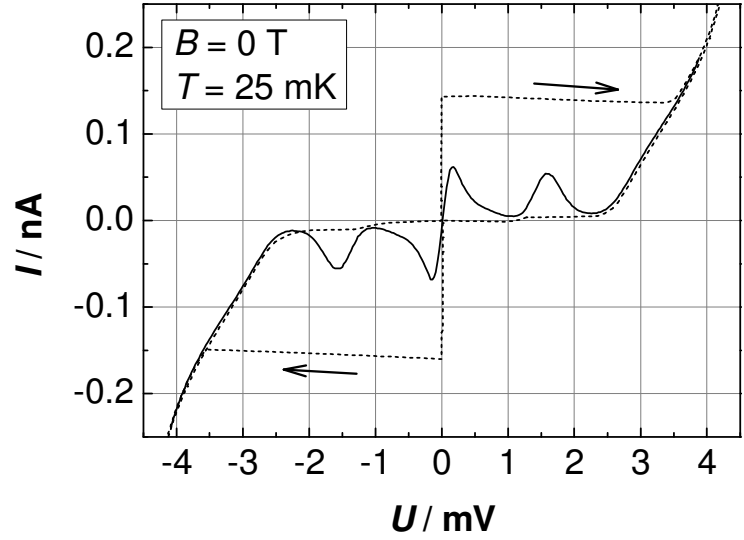

Fig. 4: *I-U* curves of transistor T1 in the superconducting state, measured in two-terminal mode, voltage biased through lines with a total resistance of 1.6 kΩ (solid), and in four-terminal configuration, current biased through a 100 MΩ resistor (dashed). The latter configuration exhibits the typical hysteretic characteristic with a switching current of about 150 pA.

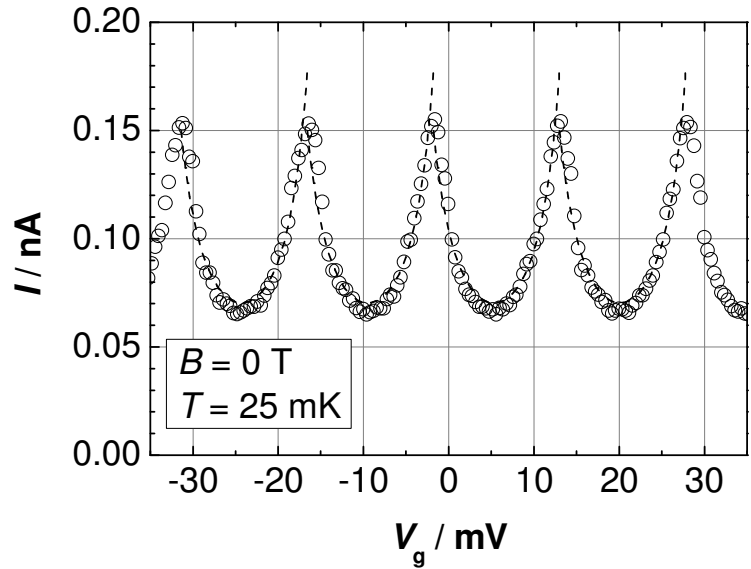

Fig. 5: Gate voltage modulation of the current through the smallest transistor (T1) in the superconducting state ($B = 0$ T), voltage biased at 140 µV (i.e. on the "supercurrent peak"). Dotted lines are theory fits $I_p \sim [E_{tr}(V_g)]^2$ according to equation (1).

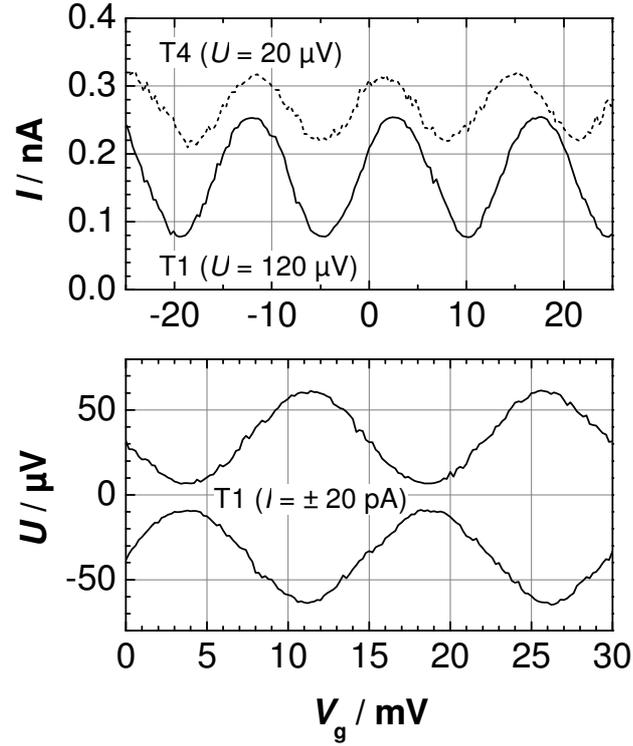

Fig. 6: Current modulation characteristics for transistor samples with the smallest (T1) and the largest (T4) junctions in the normal state ($B$ = 8 T, $T$ = 30 mK), voltage biased at their working points of maximum modulation amplitude (top panel). The bottom panel shows voltage modulation curves of T1, measured at small bias current $I = \pm 20$ pA.

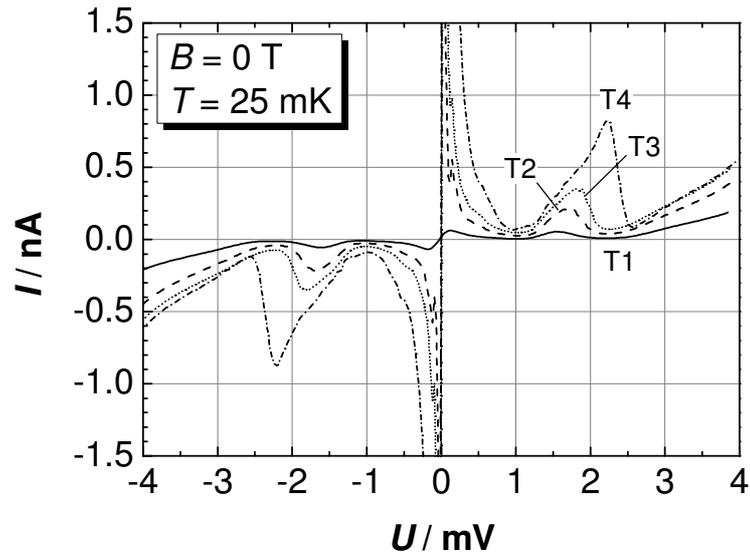

Fig. 7: Subgap regions of the $I$-$U$ curves measured on the superconducting transistor samples T1-T4 (voltage biased mode).

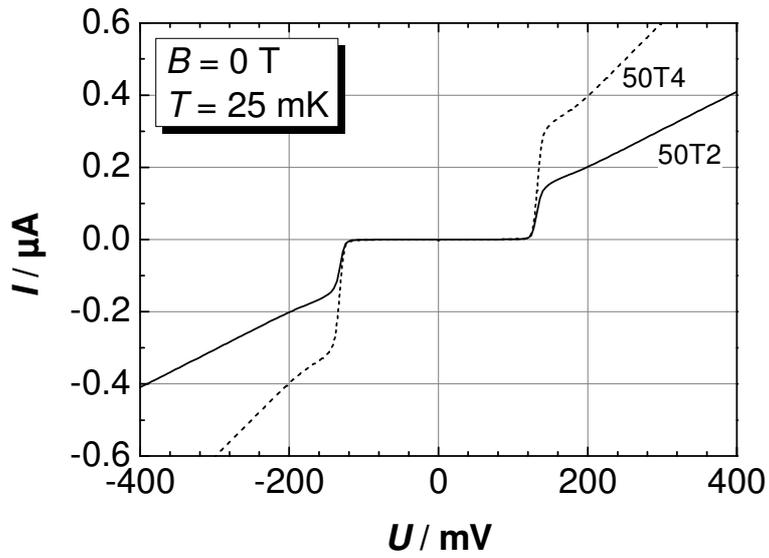

(a)

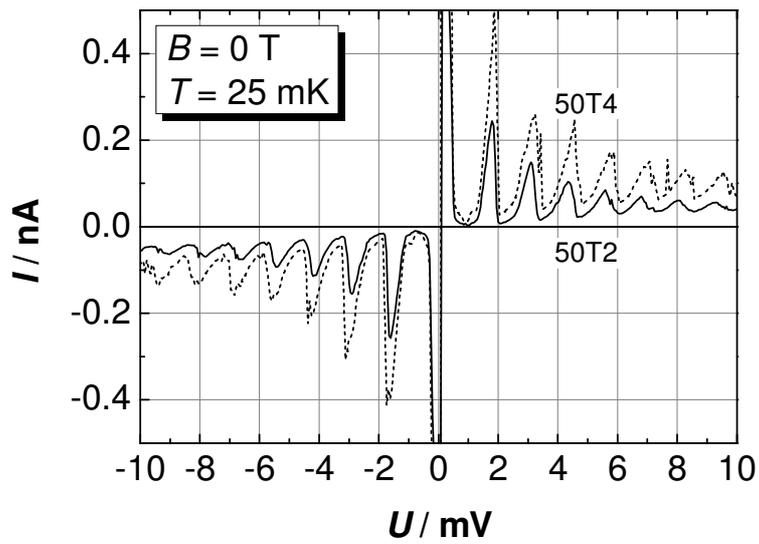

(b)

Fig. 8: *I-U* characteristics of two differently sized series array samples, each consisting of 50 junctions, in the superconducting state (a). The blowup of the subgap region (b) shows oscillatory structures, fading out with increased voltage.

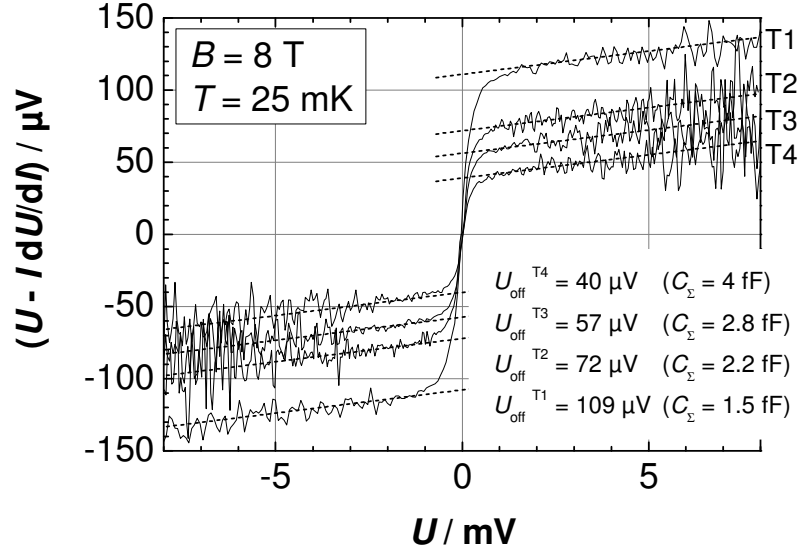

Fig. 9: Offset plot for the transistor samples T1-T4 in the completely normal state. Offset voltages $U_{\text{off}}$ were determined from the extrapolation of the linear branches to zero voltage (dotted lines), and resulting values $C_\Sigma = e/U_{\text{off}}$ are given.

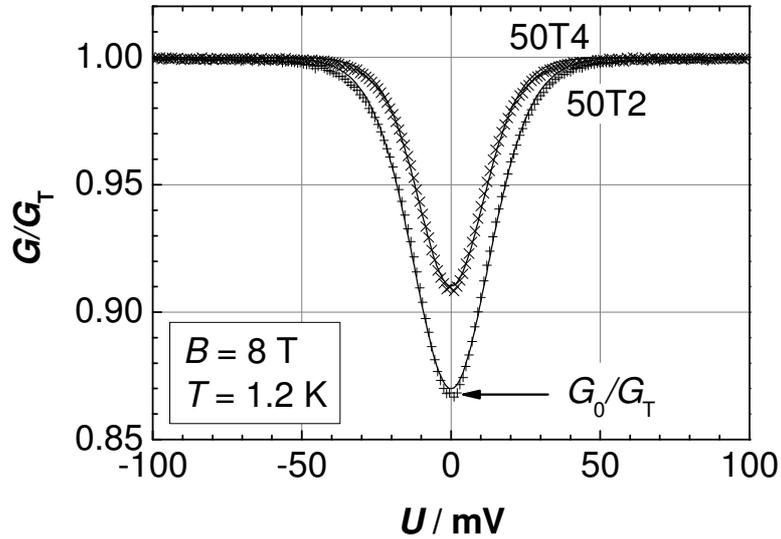

Fig. 10: Plot of the normalized conductance $G/G_T(U)$, measured for both array samples at $T = 1.2$ K (crosses). Junction capacitances $C_T = 1.9$ fF (for sample 50T2) and 2.7 fF (50T4) were calculated from $G_0/G_T$ by the CBT formula (see text). Theory fits to the data, shown as lines, yielded $C_T = 1.8$ fF (50T2) and $C_T = 2.9$ fF (50T4).

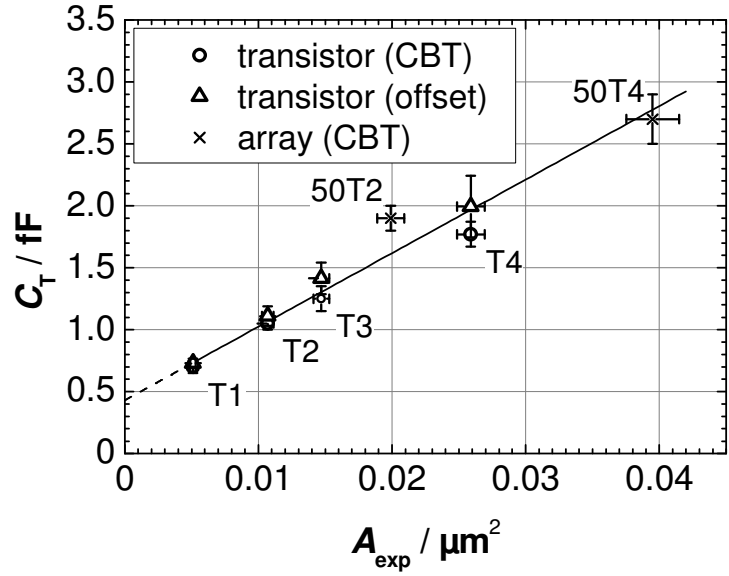

Fig. 11: Junction capacitances $C_T$ vs. junctions areas $A_{exp}$ for the transistor samples (labeled T1-T4) and for the array samples (50T2 and 50T4). For the transistors both data derived from the offset plot method (triangles) and from the CBT method (circles) are shown and coincide well. All data were fitted by the relation $C_T(A_{exp}) = A_{exp} \times 59 \text{ fF}/\mu\text{m}^2 + 0.43 \text{ fF}$ (line).